\documentclass[prc,twocolumn,superscriptaddress,showpacs,preprintnumbers,amsmath,amssymb]{revtex4}

\usepackage{gensymb}
\usepackage{graphicx}% Include figure files
\usepackage{dcolumn}% Align table columns on decimal point
\usepackage{bm}% bold math
\usepackage{color,ulem}
\newcommand{\be}{\begin{equation}}
\newcommand{\ee}{\end{equation}}
\newcommand{\bea}{\begin{eqnarray}}
\newcommand{\eea}{\end{eqnarray}}

\begin{document}

\title{Scaling of the $^{19}$B two-neutron halo  properties   close to unitarity}

\author{Emiko Hiyama}
\address{Department of Physics, Tohoku University, 980-8578, Japan}
\address{RIKEN, Nishina Center, Wako, Saitama 351-0198, Japan}

\author{Rimantas Lazauskas}
\address{Universit{\'e} de Strasbourg, CNRS, IPHC UMR 7178, F-67000 Strasbourg, France}

\author{Jaume Carbonell}
\address{Université Paris-Saclay, CNRS/IN2P3, IJCLab, 91405, Orsay, France}

\author{Tobias Frederico}
\address{Instituto Teccnol\'ogico de Aeron\'autica, 12.228-900 S\~ao Jos\'e dos Campos, Brazil}

\date{\today}

\begin{abstract}
 We explore the description of the bound $^{19}$B isotope
 in terms of a $^{17}$B+n+n three-body system where the two-body subsystems $^{17}$B+n
 and neutron-neutron (nn) have virtual states close to the continuum. Dimensionless scaling functions for the root-mean-square (rms) radii are defined and studied for different parameters of the neutron-core potential and considering three different models for neutron-neutron interaction. The scaling functions for the radii are rooted on the universal behavior of three-body systems close to the Efimov limit and  depend only on dimensionless quantities formed by the two-neutron separation energies and scattering lengths. Our results show in  practice the model independence of these scaling functions close to unitarity. We provide an estimation of the different rms relative separation distances between the constituents, as well as of the proton and matter radii.
\end{abstract}

\maketitle

\section{Introduction}

Weakly bound light radioactive nuclei having the structure of a
core+n+n  are found in the vicinity of the neutron drip line and
share the characteristic property of forming a large
two-neutron halo~\cite{NielsenPRep2001}. The neutron halo is
markedly situated outside the n-core interaction range, and it is
governed by a large n-core s-wave scattering length,  exceeding
by far the effective range. Much have been discussed on the
structure and reactions of light  two-neutron halo nuclei close to
the drip-line since the pioneering work on
$^{11}$Li~\cite{Tanihata:1985psr}, with an extensive review
literature~\cite{Jensen:2004zz,Braaten:2004rn,Meng:2005jv,Fred2012,TanihataPPNP2013,Riisager2013,Zinner:2013ema,Canto:2015esm,Naidon:2016dpf,Greene:2017cik,Hammer:2022lhx}.
To cite one example, it has been
pointed out theoretically that $^{22}$C has a n-$^{20}$C subsystem
with a large s-wave scattering length, and then as a result, the
three-body system, namely $^{20}$C-n-n, forms a  weakly-bound
Borromean state of
$^{22}$C~\cite{Mazumdar2000,22C_Yamashita_2011,22C_Phillips_2013},
which is supported by analysis of earlier
experiments~\cite{Tanaka:2010zza} and by the more recent
observation and analysis of  higher-precision  cross-sections~\cite{Togano:2016wyx}.

In 2010,  the weakly-bound Borromean state of
$^{19}$B~\cite{MSU_2012} has been observed.
The paper reported
that a virtual state of $^{18}$B
is located below the $^{17}$B+n threshold and that the scattering
length of $^{17}$B+n, $a_s$, was negative and large. However, due
to the poor resolution and acceptance of the experiment,
the precise value of the
scattering length 
was not determined and
only an upper
bound $a_s <-50$ fm was established. 
The two-neutron
separation energy, $S_{2n}$ of $^{19}$B was measured in another
experiment~\cite{B19_mass}, where it was reported to be $0.14 \pm
0.39\,$MeV. A more recent compilation of nuclear masses provides
$S_{2n}=0.09\pm 0.56$~MeV~\cite{WangChin2017}.

Motivated by these experiments, some of the present authors (E.H.,
R.L. and J.C.) studied the structure of $^{19}{\rm B}$ within the
framework of $^{17}$B+n+n three-body model~\cite{HiyamaPRC2019}
and predicted the binding energy of the ground state in $^{19}$B
by tuning the n-$^{17}$B scattering length  ($-100<a_s<-20\,$fm).
The  two-neutron separation energy value of $S_{2n}=0.13\,$MeV was found 
by considering $a_s=-100\,$fm, which is compatible with the
experimental data~\cite{B19_mass}. In addition, due to the large
magnitude of the $^{17}$B+n scattering length, it was pointed out
that the ground state has features of Efimov
state~\cite{Efimov:1970zz,Efimov:1971zz}, however due to the
unfavorable heavy-light-light mass composition of this three
cluster state it is highly unlikely to form an Efimov excited
state. That, indeed, is excluded by the possible regions of
existence of such states systematized in Ref.~\cite{AmorimPRC1997}
and later on within EFT in~\cite{Canham:2008jd}.
Despite of that, the shallow ground state of the large two-neutron
halo of $^{19}$B should exhibit  universal and model independent
properties through its structure, which would be dominated by what
is known as Efimov physics~\cite{Efimov:1990rzd}: an intrinsic
consequence of the discrete scale in-variance in the unitarity
limit,  or equivalently in the zero-range interaction limit (see
e.g.~\cite{Fred2012}).

Furthermore, a recent new experiment on $^{19}$B has been
performed~\cite{CookPRL2020}.  An enhanced soft
electric-dipole mode   just above the two-neutron decay threshold was observed,
and properly interpreted within a three-body calculations that reproduce
the energy spectrum.  They found the best fit of the relative
energy spectrum for $S_{2n}=0.5\,$MeV $a_s=-50\,$fm, resulting in
a rms radius of the core with respect the two-neutron center of
mass given by:
\begin{equation}\label{eq:exprc2n}
\sqrt{\langle r^2_{c-2n}\rangle}=5.75\pm  0.11 \,\text{(stat)}\pm
0.21\,\text{(sys)}\,\, \text{fm}\, ,
\end{equation}
similar in size to the corresponding quantity in $^{11}$Li. It was
concluded that the valence neutrons have a significant s-wave
configuration and exhibit a pronounced neutron-neutron
correlation. 
Apart from the experimental activity, 
theoretical efforts were expended to study  the  structure of
$^{19}$B in a three-body model~\cite{CasalPRC2020} and also within
the many-body mean-field approach of the deformed relativistic
Hartree-Bogoliubov theory in the continuum~\cite{Sun:2021alk}.

In present paper, 
we study the universal features of the structure of the shallow Efimov-like $^{19}$B  ground state from other perspective.
 We establish 
the  appropriate $^{19}$B scaling functions for the different rms radii close to the unitarity limit, which  depend only on two dimensionless quantities
formed by the two-neutron separation energies and scattering lengths. That approach was proposed for general three-body systems interacting  via 
s-wave zero-range potentials in Ref.~\cite{Yamashita:2004pv}. Such scaling functions describe 
universal correlations between observables and they appear as limit-cycles from the discrete scale symmetry that the system presents when the potential range is driven to zero or the scattering lengths to infinity~\cite{Fred2012}, which can also be built within the context of  EFT~\cite{Canham:2008jd}. The limit-cycles for the scaling functions associated to   correlations between dimensionless  quantities are built from successive Efimov or Thomas collapsed states.  

These correlations between observables represent, modulo effective-range corrections, the results obtained from short-range potentials for  shallow states.  They should be  quite model independent, which can be verified with the use of different potentials that share 
the same two-body  low-energy s-wave observables. In the light of the concepts of universal scaling functions and using the  newly extracted 2n separation energy, $a_s$, and $\langle r^2_{c-2n}\rangle^\frac12$ for this system~\cite{CookPRL2020}, we compute the $^{19}$B matter and proton rms radii, and also the different rms relative separation distances. We    check the consistence of these extracted data 
comparing results of different potential models for the n-core and nn systems, as well as,  calculations from other authors, to assert the model independence of the scaling laws.  Therefore, we expect that this work can be also viewed as an useful model independent systematic to predict long wavelength observables on the basis of a minimal number of physical inputs.

The work is organized as follows. In Sect.~\ref{sect:model}, we  briefly review the  model from Ref.~\cite{HiyamaPRC2019} for the neutron-core interaction based on a parametrization  of the MT13 potential. In Sect.~\ref{sect:results}, we introduce the relevant scaling functions for the two-neutron separation energy and the different rms radii. In the sequence, we present the calculation of the scaling function for $S_{2n}$, the 
radii vs. $a_s$ and $S_{2n}$ and
perform the scaling analysis for the rms radii, aiming 
to demonstrate the model independence of this correlation. In Sect.~\ref{sect:finalremarks} our final remarks are presented; we show in addition some results for the matter and proton rms radius based on our analysis through the scaling functions.

\section{Modeling $^{19}$B} \label{sect:model}

As it was proposed in Ref.~\cite{HiyamaPRC2019}, the $^{19}$B isotope is described as a $^{17}$B+n+n three-body system.
To this aim, we have first constructed an effective low-energy n-$^{17}$B potential,
simulating the short-range Pauli repulsion at distances smaller than the 
 $^{17}$B core radius, which is outweighed by the folded attractive nucleon-nucleon interaction beyond  this overlap region. 
 A simple form accounting for these facts is:
\begin{equation}  \label{V}
  V_{n^{17}{\text{B}}}(r) \ = \ V_r \, \frac{e^{-2\mu r}}{r}\,\left[ 1 - e^{-\mu (R-r)} \right] \, ,
\end{equation}
 where $R$ is the hard-core radius
 and $\mu$ is a range parameter for the Yukawa $n$-$^{17}$B potential.
 We have chosen the value $R=3$~fm, which corresponds to the rms matter radius of $^{17}$B~\cite{B17-19_RMS},
 and we take $\mu=0.7$~fm$^{-1}$ corresponding to the pion mass.
 
\begin{figure}[h!]
\vspace{0.5cm}
\begin{center}
\includegraphics[angle=0,width=0.36\textwidth]{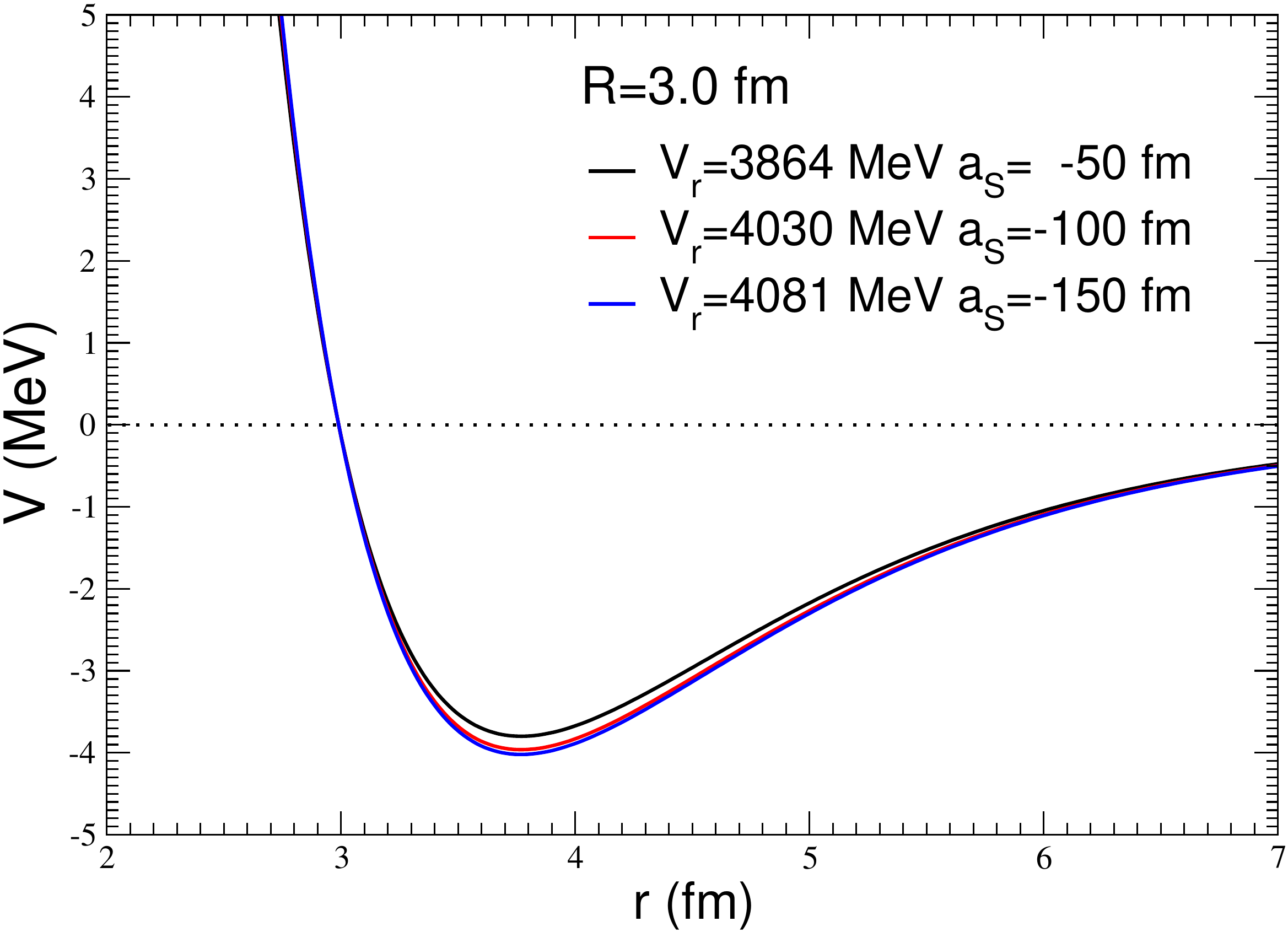}
\caption{$V_{n^{17}B}$ potential (MeV) reproducing several values  of the n-$^{17}$B scattering length $a_s$ (fm) 
and the corresponding strength parameters $V_r$}\label{V_n17B}
\end{center}
\end{figure}

 Once fixed the range $\mu$ and the size $R$, potential (\ref{V}) depends on a single strength parameter $V_r$,
 which is 
 tuned to reproduce the n-$^{17}$B scattering length $a_S$.
 We display in Fig.  \ref{V_n17B} the potentials reproducing the values $a_S=-50,-100,-150$ fm
 with the corresponding strength parameters $V_r$ (MeV).

 All the numerical values along this work correspond to $m_n=939.5654$~MeV, $m_{^{17}{\text{B}}}=15879.1$~MeV, 
 i.e.\ a $n$-$^{17}$B reduced mass $m_R=887.0771$~MeV and ${\hbar^2/2m_R}=21.9473$~MeV$\cdot$fm$^2$.

The interaction \eqref{V} is purely central and independent on  the total spin $\vec{S}$ 
of the n+$^{17}$B system ($\vec{S}$=$\vec{s}_n$ + $\vec{S}_{^{17}B}$).
Since $^{17}$B is a $J^{\pi}=3/2^-$ state, it can couple to a neutron 
in two different total spin states, $S=1^-,2^-$.
This "spin-symmetric" approximation, could be not very realistic
given that, according to \cite{MSU_2012}, the  $S=2^-$ state
has a resonant scattering length and there is
no reason that the $S=1^-$ would be resonant as well.
Nevertheless in sake of simplicity we decided to stick with a 
spin-independent form of  n+$^{17}$B interaction, whose strength is
directly associated with a single 
unknown:  n+$^{17}$B scattering
length $a_S$.

The $^{19}$B Hamiltonian requires also  the nn interaction
and we have used in our calculations three different $V_{nn}$ models. 
We have considered the charge independent (CI)  Bonn~A model~\cite{Bonn_A}
providing the nn low energy parameters  $a_{nn}$=-23.75 fm and $r_{nn}$=2.77 fm, 
and the charge dependent (CD) AV18~\cite{Wiringa:1994wb}
for which $a_{nn}$=-18.8 fm and $r_{nn}$=2.83 fm, both models acting in all partial waves. 
We have also considered the S-wave CD version of MT13~\cite{MT_NPA127_1969} built in~\cite{HiyamaPRC2019} 
with low-energy parameters $a_{nn}$=-18.59 fm and $r_{nn}$=2.93 fm, in agreement with the experimental values. 
It takes the form
 \begin{equation}\label{V_nn}
 V_{nn} \ = \ V_R\, \frac{e^{-\mu_R r}}{r} \, - \, V_A\, \frac{e^{-\mu_A r}}{r} \, ,
\end{equation}
where the parameters are: $V_R$=1438.720 MeV$\cdot$fm, $V_A$=509.40 MeV$\cdot$fm, $\mu_R^ {-1}$=3.11 fm, $\mu_A^ {-1}$=1.55 fm. 

\section{$^{19}$B results}
\label{sect:results}

 We have  computed the $^{19}$B ground-state energy, measured by the two-neutron separation energy $S_{2n}$ as a function of
 $a_s$. The rms relative separation distances were also calculated for $\langle  r^2_{nn}\rangle^\frac12$,
 $\langle  r^2_{nc}\rangle^\frac12$, $\langle  r^2_{c-2n}\rangle^\frac12$, $\langle  r^2_{n}\rangle^\frac12$ and
 $\langle  r^2_{c}\rangle^\frac12$. The three-body problem was solved 
 by the Gaussian Expansion Method~\cite{Hiya03} and by the Faddeev equation formalism in configuration space~\cite{LC_3n_PRC72_2005,FE_3B_20054}.
 
 In Table~\ref{tab:results}, we present an example of our results  obtained with both Bonn-A and MT13 potentials, where the dependence on the 
 spin independent
 n+$^{17}$B scattering length is explored for the two-neutron separation energy and  different rms relative separation distances in the Borromean $^{17}$B+n+n system.
 As it should be, $S_{2n}$ increases towards unitarity, as $V_{n^{17}B}$ becomes slightly more attractive, despite that the two neutron separation energy is below 0.2~MeV, implying in a giant halo around 10~fm to be compared with the smaller size of the core nucleus. Such values of $S_{2n}$ are  consistent with $S_{2n}=0.14\pm0.39$~MeV~\cite{B19_mass}. 
 
 Furthermore, by using the results presented in Table~\ref{tab:results} the neutron and core rms  distances to the center of mass can be easily evaluated from $\langle r^2_{nn} \rangle^\frac12$, $\langle r^2_{c-2n} \rangle^\frac12$ and $\langle r^2_{nc} \rangle^\frac12$, given respectively by:
 \begin{equation}
  \langle r^2_{n} \rangle^\frac12=   \sqrt{\frac{2A\,\langle r^2_{nc} \rangle + \langle r^2_{nn} \rangle}{2(A+2)} - \frac{2A\,\langle r^2_{c-2n} \rangle}{(A + 2)^2} }\, ,
 \end{equation}
 and $\langle r^2_{c} \rangle^\frac12 = 2 \langle r^2_{c-2n} \rangle^\frac12/(A + 2)$, with $A$ the  $^{17}$B mass number. Also, the average relative angles, which are given by:
\begin{eqnarray}
&&\theta_{nc} =\cos^{-1}\frac{\langle \vec r_{c}\cdot \vec r_{n}\rangle}{\sqrt{\langle r^2_{c} \rangle \langle r^2_{n} \rangle}}=\cos^{-1}  \frac{\langle r^2_{c} \rangle + \langle r^2_{n}\rangle - \langle r^2_{nc} \rangle}{2\sqrt{\langle r^2_{c} \rangle \langle r^2_{n} \rangle}}\, , \nonumber
\\ &&\theta_{nn'}=
\cos^{-1}\frac{\langle \vec r_{n}\cdot \vec r_{n'}\rangle}{ \langle r^2_{n} \rangle} =\cos^{-1} \left( 1- 
\frac12 \frac{\langle r^2_{nn}\rangle}{\langle r^2_{n} \rangle}\right)\, , \label{eq:angles}
\end{eqnarray}
 quantifies the geometry of the halo and verifies
 $$ \theta_{nn'}+2\,\theta_{nc}=360^{\circ}\, ,$$
 that is accurately fulfilled in our calculations.

\begin{table}[thb!]
     \centering
          \caption{Dependence on the singlet n+$^{17}$B scattering length of the two-neutron separation energy and rms relative separation distances in the $^{17}$B+n+n system computed with the Bonn-A and MT13 $nn$ potentials for $R=3\,$fm. $S_{2n}$ in MeV,  $a_s$ and  $\langle r^2_{\alpha} \rangle^\frac12$ in fm $(\alpha=nn,\,c-2n,\,nc)$.}
     \label{tab:results}
     \begin{tabular}{ccccc}
     \hline
Bonn-A   \\
\hline
$a_s$ & $\qquad S_{2n}\qquad$ & $~\langle r^2_{nn} \rangle^\frac12~$& $~\langle r^2_{c-2n} \rangle^\frac12~$& $~\langle r^2_{nc} \rangle^\frac12~$\\
\hline
-50 &     0.087031  &  16.15 &   10.89  & 13.56   \\ 
-80  &     0.117391 &  14.79 &    10.11  &    12.53 \\    
-100 &    0.128790  &  14.40 &   9.89  &   12.23   \\ 
-300 &    0.162571 &    13.45 &  9.35   &  11.52  \\  
-500 &    0.169934 &    13.28 &   9.25  &   11.39 \\   
-1000 &  0.175789  &   13.15  &    9.18   &    11.29  \\    
\hline
MT13\\
\hline
$a_s$ & $S_{2n}$ & $\langle r^2_{nn} \rangle^\frac12$& $\langle r^2_{c-2n} \rangle^\frac12$& $\langle r^2_{nc} \rangle^\frac12$\\
\hline
-50  &    0.063234  &    18.24  &  11.78  &   14.90 \\
-80  &    0.090628  &    16.43  &  10.77  &  13.55  \\
-100 &    0.101013  &    15.92  &   10.49 &   13.17  \\
-300 &    0.132020  &    14.72  &   9.82 &   12.27 \\
-500 &    0.138816  &    14.51  &   9.70 &   12.11 \\
-1000 &   0.144228  &   14.35   &   9.61 &   11.99 \\
\hline
     \end{tabular}
 \end{table}

%%%%%%%%%%%%%%%%%%%%%%%%%%%%%%%%
 \subsection{Scaling functions}
 
Within the considered model, the $^{17}$B+n+n system is loosely bound, such that its giant halo has rms relative distances by one order of magnitude 
larger than the interaction range $\sim 1$~fm. One should notice however that the $nn$ effective ranges are only about four times smaller than these sizes,
 thus indicating a possible relevance of the range corrections.
 Under this condition we will study the results of Table~\ref{tab:results} systematically employing the halo universal scaling laws (see e.g.~\cite{Fred2012}), emerging as a consequence  of the correlations between observables of Efimov-like states. 
 
 Close to the unitarity limit, these scaling laws depend only on the scattering lengths, the two neutron separation energy and core mass number. The  energy scaling function as introduced in~\cite{MadeiraPRA2021},  is the correlation between the three-body  energy at a given scattering length with  the value at the unitarity. The range dependence was also taken into account, which we will not display here:
\begin{eqnarray}
\frac{S_{2n}}{S^{un}_{2n}}=\mathcal{F}\left(\left[a_s \kappa_{nc}\right]^{-1}\hspace{-.1cm},\left[a_{nn}\kappa_{nn}\right]^{-1}\hspace{-.1cm},A\right) ,\label{eq:s2n}
\end{eqnarray}
where  
\begin{equation}
\kappa_{nc}=(2\mu_{nc}S_{2n})^\frac12,\quad \kappa_{nn}=(m_nS_{2n})^\frac12\, ,
\end{equation}
and $\mu_{nc}$ is the reduced mass of the n-core system 
 and $S^{un}_{2n}$ is the two-neutron separation energy at unitarity. The last scaling relation is written by considering $\hbar=m_n=1$ and
it attains the trivial value 1 at unitarity.
 
 Other scaling functions can be found for  the rms relative separation distances, and they read~\cite{Yamashita:2004pv} (see also~\cite{Fred2012} for further discussions):
 \begin{equation}
 \langle r^2_\alpha \rangle\sqrt{S_{2n}}= \mathcal{R}_\alpha\left(\left[a_s \kappa_{nc}\right]^{-1}\hspace{-.1cm},\left[a_{nn}\kappa_{nn}\right]^{-1}\hspace{-.1cm}, A \right)\, ,
 \label{eq:scalr1}
 \end{equation}
where $\alpha$ denotes the possible relative distances: $nn$ (neutron-neutron), $nc$ (neutron-core), $c-2n$ (core - $nn$ c.m.), $n$ (neutron-c.m.) and $c$ ($^{17}$B-c.m.). The angles defined by Eq.~(\ref{eq:angles}) are as a consequence of Eq.~(\ref{eq:scalr1}) scaling functions determined, in principle, by the limit-cycles of the different radii.
 
 \begin{figure}[t]
\begin{center}

\includegraphics[angle=0,width=0.46\textwidth]{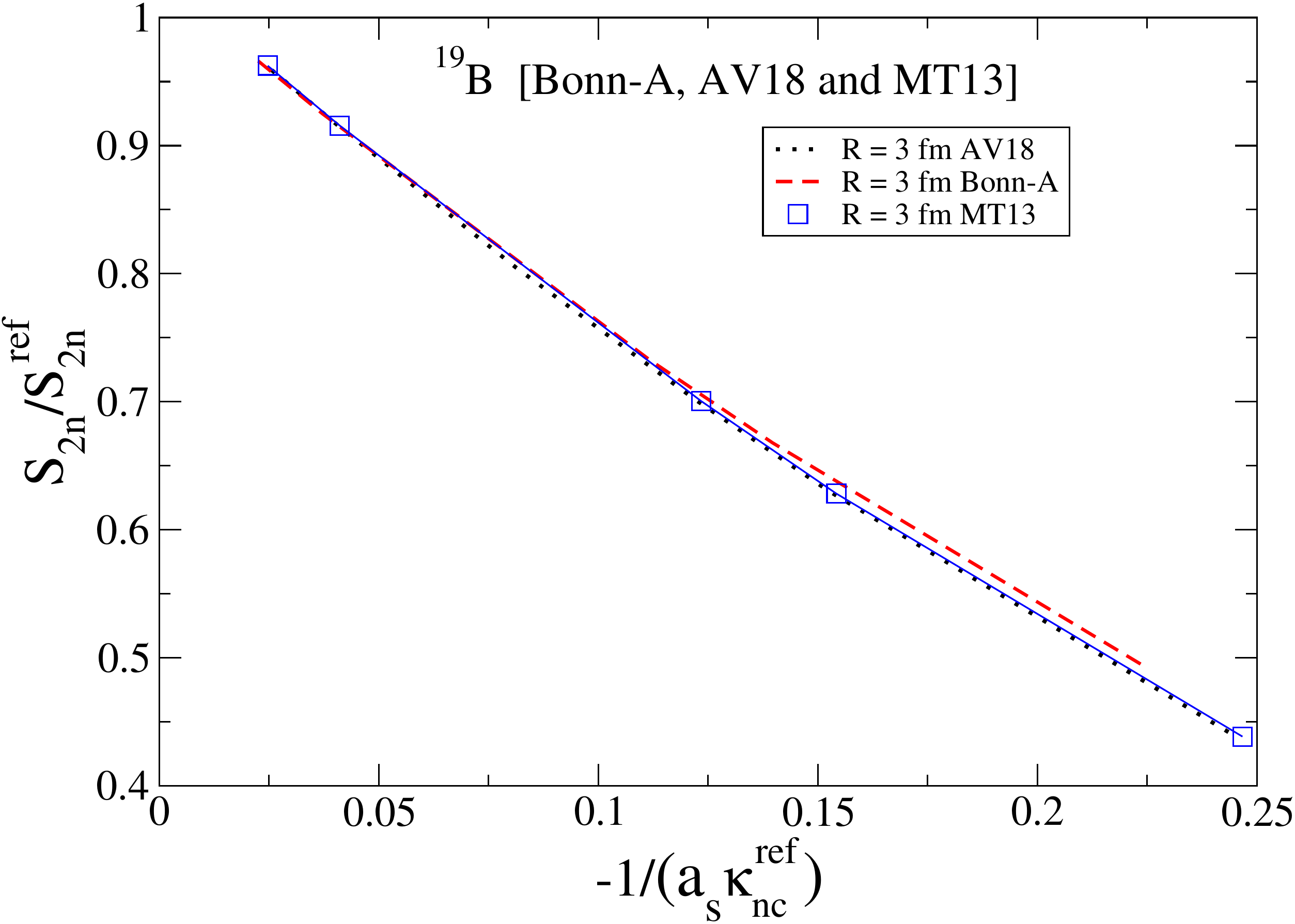}

\caption{$^{19}$B ratio of two-neutron separation 
and a reference one, $S_{2n}/S^{\text{ref}}_{2n}$ as 
function of $-1/(2\mu_{nc}a^2_s S^{\text{ref}}_{2n})^\frac12$. The  results are obtained with $R=3\,$fm for the 
neutron-neutron Bonn-A potential (dashed line) and   for MT13-potential (empty squares). The calculations for AV18  were performed with $R=\,3\,$fm (dotted line). The reference two-neutron separation energy is 
explained in the text. }\label{Fig:19B-Bonn-MT-5}
\end{center}
\end{figure}
 
The unitarity limit of the scaling relations are found for $1/(a_s \kappa_{nc})=1/(a_{nn}\kappa_{nn})=0$ written as:
 \begin{equation}
 \langle r^2_\alpha \rangle_{un}\sqrt{S^{un}_{2n}}= \mathcal{R}_\alpha\left(0,0, A \right)\, ,
 \label{eq:scalr2}
 \end{equation}
 which was discussed in~\cite{Fred2012}, and being only a function of the core mass number.  Another universal property of the scaling function (\ref{eq:scalr1}) is~\cite{Fred2012}:
 \begin{eqnarray}
   \frac{\partial}{\partial z_i} \mathcal{R}_\alpha\left(z_1,z_2, A \right)>0  \, , 
   \label{eq:derivative}
 \end{eqnarray}
which means, in particular, that the halo system 
shrinks when it moves from unitarity to a Borromean configuration. This is natural as the system has to 
compress in order to preserve the same binding energy when it is driven from unitarity to a Borromean state~\cite{Yamashita:2004pv}. Indeed, such behavior is also confirmed in present calculations, as we will illustrate in what follows.

%%%%%%%%%%%%%%%%%%%%%%%%%%%%%%%%%%%%%%%%%%%%%
\subsection{Scaling analysis for $S_{2n}$}

The scaling function of the two-neutron separation energy  for the Borromean system
$^{19}$B  is shown in Fig.~\ref{Fig:19B-Bonn-MT-5} in terms of a reference value:
    \begin{eqnarray}
\frac{S_{2n}}{S^{\text{ref}}_{2n}}=\mathcal{F}\left(\left[a_s \kappa^{\text{ref}}_{nc}\right]^{-1}\hspace{-.1cm},\left[a_{nn}\kappa^{\text{ref}}_{nn}\right]^{-1} ,A\right),\label{eq:s2nref}
\end{eqnarray}
with the definitions:
\begin{eqnarray}
\kappa_{nc}^{\text{ref}}=\left(2\mu_{nc}S^{\text{ref}}_{2n}\right)^{\frac12}\, \text{and}\,\,
\kappa_{nn}^{\text{ref}}=\left(m_{n}S^{\text{ref}}_{2n}\right)^{\frac12}.~~
\end{eqnarray}
We make use of the general form provided by Eq.~(\ref{eq:s2n}), with the arguments corresponding to   $a_s=-1000\,$fm for $S^{\text{ref}}_{2n}=0.1758\,$MeV and $S^{\text{ref}}_{2n}=0.1442\,$MeV, as given in Table~\ref{tab:results} for Bonn-A and MT13 nn potentials, respectively. For the AV18 interaction we have $S^{\text{ref}}_{2n}=0.1481\,$MeV for $R=3\,$fm and $a_s=-1000\,$fm.

We observe that the slope is weakly dependent on the nn interaction model, confirming the validity of an universal scaling law. The curves in the figure can be parameterized by:
\begin{eqnarray}
\frac{S_{2n}}{S^{\text{ref}}_{2n}}\Big|_{Bonn-A}=1.03-\frac{2.94}{a_s\kappa^\text{ref}_{nc}} +\frac{2.43}{(a_s\kappa^\text{ref}_{nc})^2}+
\cdots\, ,\nonumber
% BonnA 1.03095	-2.94093 2.432
\\
 \frac{S_{2n}}{S^{\text{ref}}_{2n}}\Big|_{MT13}=1.04
 -\frac{3.00}{a_s\kappa^\text{ref}_{nc}} +\frac{2.38}{(a_s\kappa^\text{ref}_{nc})^2}+\cdots \, ,
 \nonumber
%MT13  1.03509	-3.00476 2.37618
\\
 \frac{S_{2n}}{S^{\text{ref}}_{2n}}\Big|_{AV18}=1.04
 -\frac{3.02}{a_s\kappa^\text{ref}_{nc}} +\frac{2.42}{(a_s\kappa^\text{ref}_{nc})^2}+\cdots \, ,
 \hspace{.1cm} \label{eq:params2n}
%AV18 1.03468 -3.02331 2.42113
\end{eqnarray}
which also shows that the difference between the nn scattering lengths and effective ranges of Bonn-A and MT13 is not significant for this particular scaling function. The 3-4\% deviation from unity at $1/a_s=0$ in Eq.~(\ref{eq:params2n}) shows that the reference $S^{\text{ref}}_{2n}$ value are indeed  quite close to this situation. As a reference the linear coefficients, associated with the Tan's contact~\cite{TanAnnPhys2008} are close to 3 in the three cases, which could be compared to 2.11~\cite{AmorimPRC1992,CastinPRA2011,MadeiraPRA2021} for the three-boson system.  The quadratic coefficients are about 2.4 compared to 0.80 in the three-boson case~\cite{,MadeiraPRA2021}. We observe that this correlation is quite insensitive to  the different  values of $a_{nn}$ in the three potential models for a fixed $R$. Furthermore, as the system  
shifts from the unitarity to Borromean, the separation energy tends to decrease with respect to the unitarity value, since the interaction becomes weaker.

\subsection{Radii vs. $a_s$ and $S_{2n}$}

\begin{figure}[t]
\begin{center}

\includegraphics[angle=0,width=0.46\textwidth]{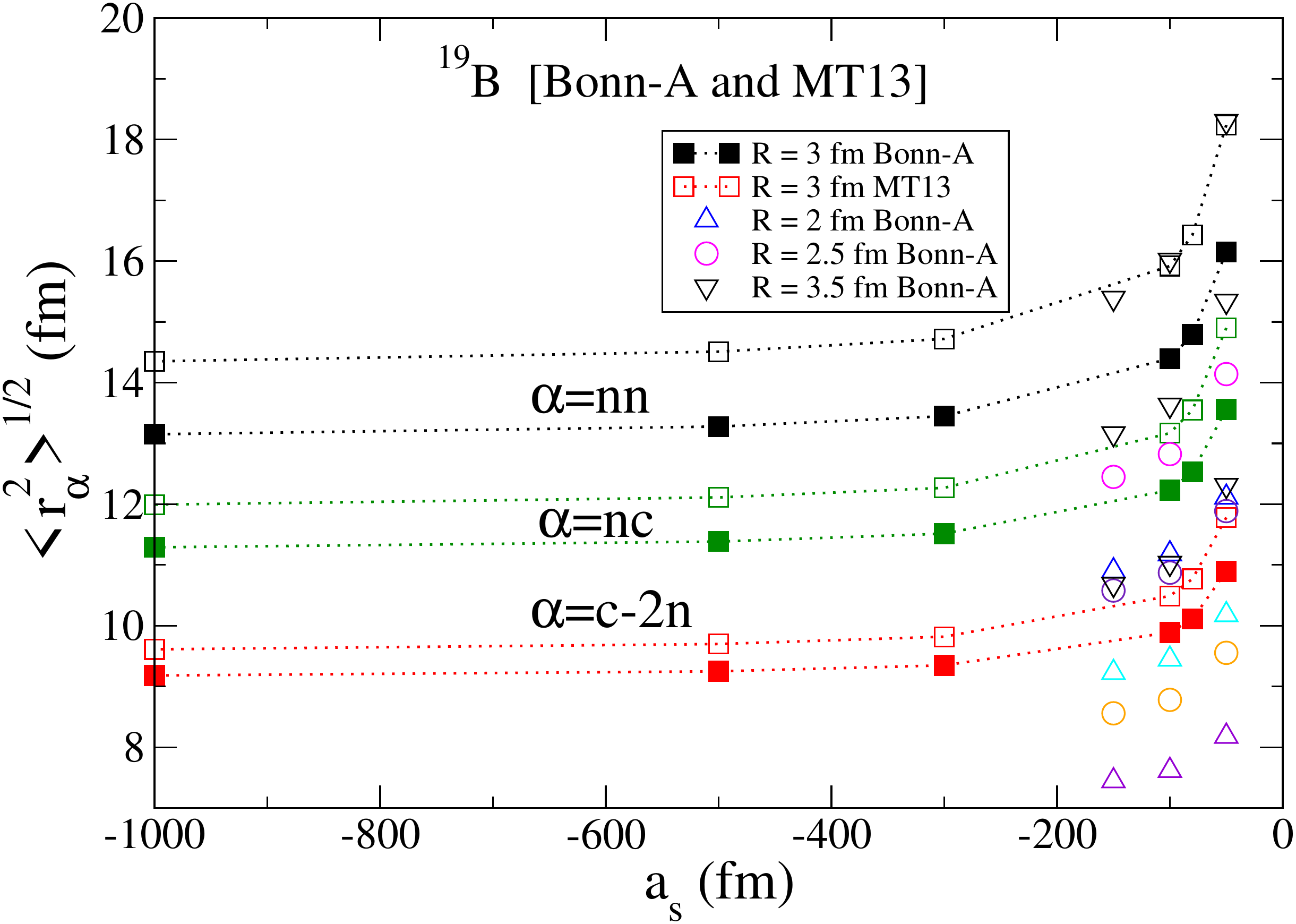}

\caption{$^{19}$B rms radii $\langle r^2_\alpha\rangle ^\frac12$ $(\alpha\equiv nn,\, nc\, , c-2n)$ as a function of the singlet scattering length. The full squares are the results for neutron-neutron Bonn-A potential and empty squares are for MT13-potential with $R=3\,$fm. The dashed lines connect the results.}\label{Fig:19B-Bonn+MT-3}
\end{center}
\end{figure}

The $^{19}$B r.m.s. radii $\langle r^2_\alpha\rangle ^\frac12$ $(\alpha\equiv nn,\, nc\, , c-2n)$  as a function of $a_s$ are displayed in Fig.~\ref{Fig:19B-Bonn+MT-3}, for the calculations with $R=3\,$ fm, given
in Table~\ref{tab:results}, and also for $R$ assuming the values of 2, 2.5 and 3.5 fm in the case of the Bonn-A potential.
The r.m.s. radii for the Bonn-A potential are systematically smaller than the ones for MT13. The size of $^{19}$B
increases when the scattering lengths decrease, both for the dependence in $a_s$ and $a_{nn}$. 

\begin{figure}[t]
\begin{center}
\includegraphics[angle=0,width=0.46\textwidth]{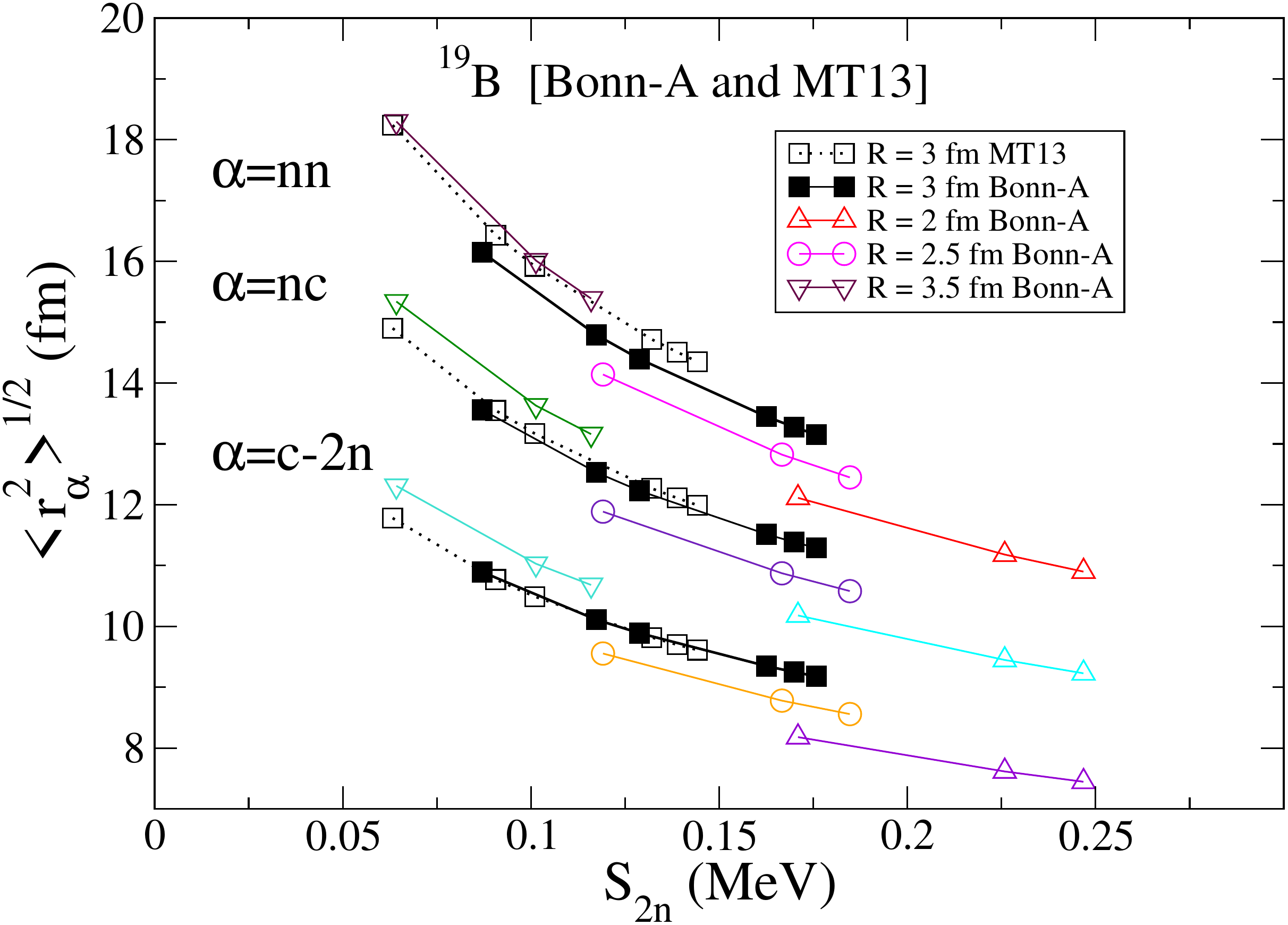}
\caption{$^{19}$B root mean square radii $\langle r^2_\alpha\rangle ^\frac12$ $(\alpha\equiv nn,\, nc\, , c-2n)$ as a function of the two-neutron separation energy. The full squares are the results for  the neutron-neutron Bonn-A potential and $R=3\,$fm. The empty squares are results for the MT13-potential obtained with $R=3\,$fm. The  lines are added just to guide an eye.}\label{Fig:19B-Bonn+MT-4}
\end{center}
\end{figure}

At a first sight, it seems 
to be in conflict with  the monotonic behavior found for the r.m.s. radii scaling functions expressed by Eq.~(\ref{eq:derivative}). 
However, we must remind the reader that also the correlation with $S_{2n}$ is relevant for these $^{19}$B quantities, as the three-body system in this lowest angular momentum state is sensitive to the short range physics determining the actual values of the separation energies. 
This feature is revealed in
Fig.~\ref{Fig:19B-Bonn+MT-4} where the results for the different radii from Table~\ref{tab:results}
demonstrate a strong correlation with the two-neutron separation energy. However, the values of the dimensionless products $1/(a_s\kappa_{nc})$ and
$1/(a_{nn}\kappa_{nn})$
move along these curves, as well as the different values of the core radius potential parameter $R$, which takes the values  2, 2.5, 3 and 3.5 fm, making the results somewhat scattered in the figure. 

In order to organize these results, it is convenient to study the radii scaling laws for the three-body system in more detail.

%%%%%%%%%%%%%%%%%%%%%%%%%%%%%%%%%%%%%%%%%%%%%%
\subsection{Scaling analysis for the rms radii}

Our task now is to provide the scaling analysis of the dimensionless products $(\langle r_\alpha^2\rangle S_{2n})^\frac12$ $(\alpha\equiv nn\, , nc\, ,c-2n)$, neutron-c.m. $(\alpha=n)$ and core-cm $(\alpha=c)$ as a function of $-1/(a_s\kappa_{nc})$ for the three employed  neutron-neutron potentials: Bonn-A, AV18 and MT13. In the present analysis the scaling functions, as formulated in Eq.~(\ref{eq:scalr1}), are plotted, in what follows, against $1/(a_s\kappa_{nc})$.  
Here one should notice that the products $1/(a_{nn}\kappa_{nn})$ are implicitly running with $S_{2n}$. 

\begin{figure}[t]
\begin{center}

\includegraphics[angle=0,width=0.46\textwidth]{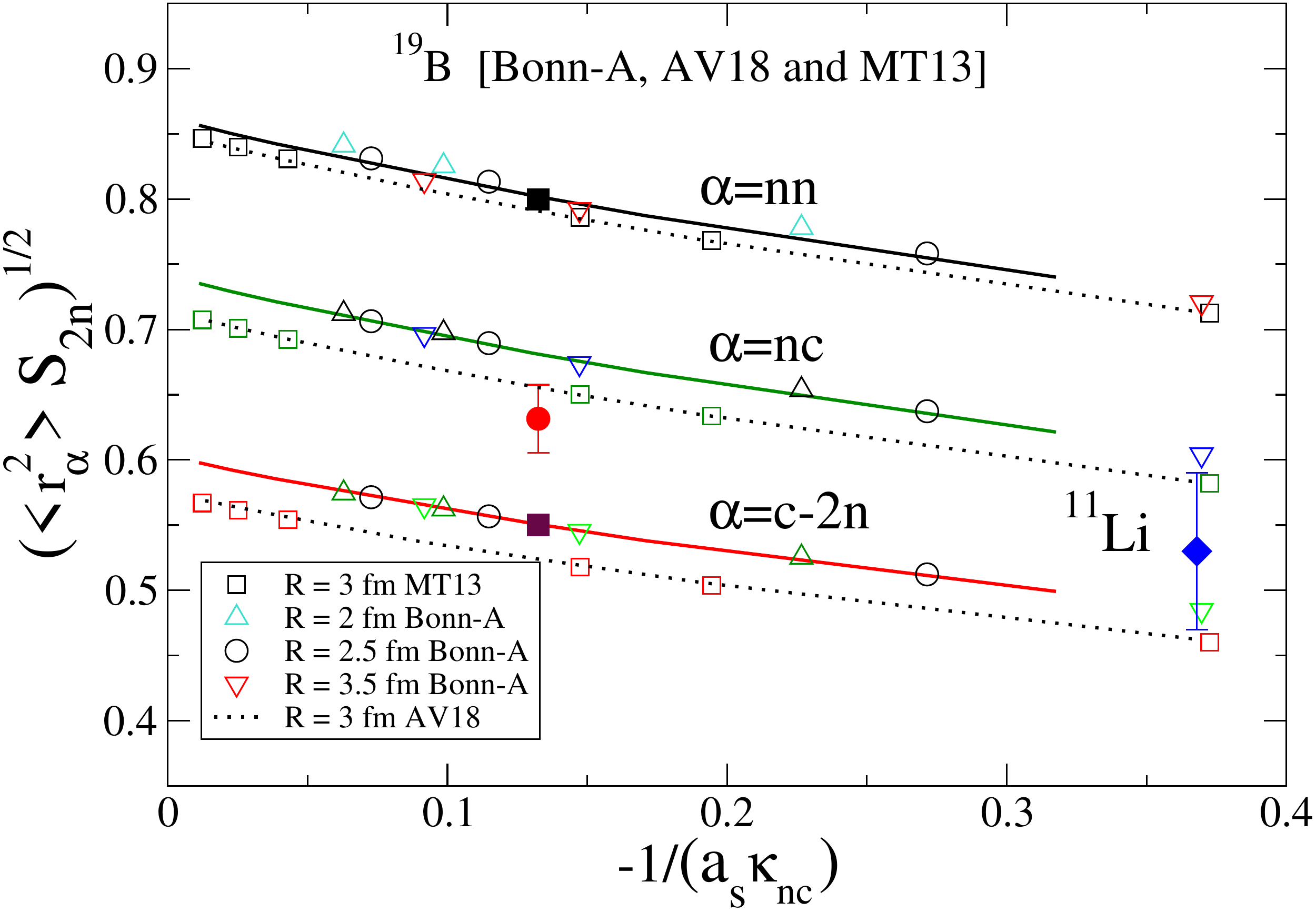}

\caption{$^{19}$B dimensionless products $(\langle r_\alpha^2\rangle S_{2n})^\frac12$ $(\alpha\equiv nn\, , nc\, ,c-2n)$  as a function of $-1/(a_s\kappa_{nc})$, where $\kappa_{nc}=(2\mu_{nc}S_{2n})^\frac12$. The solid, long-dashed and dashed lines are the results for Bonn-A potential and empty squares  for MT13-potential with $R=3\,$fm. The calculations with $R$ equals to 2, 2.5, and 3.5 fm obtained with Bonn-A potential are the up-triangle, circle and down-triangle, respectively. The results for AV18  for $R=3\,$fm are shown by the dotted line.
The full squares corresponds to  $(\langle r^2_{c-2n}\rangle  S_{2n})^\frac12=0.55$ and $(\langle r^2_{nn}\rangle  S_{2n})^\frac12=0.80$ for $1/(a_s\kappa_{nc})=-0.1325$
from  the three-body calculations of Casal and Garrido~\cite{CasalPRC2020}. The full circle is the experimental extraction of Ref.~\cite{CookPRL2020} with values $(\langle r^2_{c-2n}\rangle  S_{2n})^\frac12=0.632\pm 0.026$ and $1/(a_s\kappa_{nc})=-0.1325$. The full diamond is the estimated value of $(\langle^{11}\text{Li}| r^2_{c-2n}|^{11}\text{Li}\rangle  S^{^{11}\text{Li}}_{2n})^\frac12$ (explanation in the text).}\label{Fig:19B-Bonn+MT-1}
\end{center}
\end{figure}

 In Fig.~\ref{Fig:19B-Bonn+MT-1} the dimensionless products $(\langle r_\alpha^2\rangle S_{2n})^\frac12$ $(\alpha\equiv nn\, , nc\, ,c-2n)$  as a function of $-1/(a_s\kappa_{nc})$, for both Bonn-A, AV18 and MT13 potentials are presented. 
 The zero-range limit at unitarity  $(\langle r_{nn}^2\rangle S_{2n})^\frac12|_{zr} =0.8$ and $(\langle r_{nc}^2\rangle S_{2n})^\frac12|_{zr} =0.7$ (see ~\cite{Fred2012} and references therein) are reasonable close to the present calculations when  $1/(a_s\kappa_{nc})$ 
 approaches zero. It is worth noticing that both the neutron-neutron scattering lengths as well as effective ranges are finite, which of course, differentiate from the strict unitarity limit, defined by infinite scattering lengths and vanishing effective ranges. 
 
 The effect of the  $a_{nn}$, being slightly different for Bonn-A and MT13 or AV18, might be noticed in the figure, with the MT13 and AV18 results systematically lower than the Bonn-A ones. On the other hand,  $a_{nn}$ for AV18 and MT13 differ only by 1\% and this difference is not enough to make 
 an observable shift in the figure. %
 The somewhat lower values obtained for MT13 and AV18 with respect to Bonn-A, is due to 
 $$1/(a_{nn}\kappa_{nn})^\frac12|_{MT13}<1/(a_{nn}\kappa_{nn})^\frac12|_{Bonn-A}\, ,$$
 which turns the two-neutron halo of $^{19}B$ more compact for the MT13 model,  consistent with the monotonically increasing behavior expressed by the partial derivatives given in Eq.~(\ref{eq:derivative}) for the radii scaling functions.

 We compare the scaling functions in Fig.~\ref{Fig:19B-Bonn+MT-1} with the calculation from~\cite{CasalPRC2020}. The last calculations employed a Woods-Saxon (WS) plus spin-orbit neutron-$^{17}$B potential, the Gogny-Pires-Tourreil (GPT) neutron-neutron one~\cite{GognyPLB1970} ($a_{nn} =-22.12\,$fm and $r_{nn}=2.83\,$fm) in conjunction with hyperradial Gaussian three-body potential tuned to fit $S_{2n}=0.5\,$MeV. The central part of the WS potential gives $a_s=-50\,$fm, and the strength of the spin-orbit term is determined to fit the position of a $d_{5/2}$ resonance at 1.1~MeV above the $^{17}$B-neutron continuum close to the $1^-$ state from shell-model calculations~\cite{MSU_2012}. Their results are
 $\langle r^2_{nn} \rangle^\frac12= 7.28\,$fm and $\langle r^2_{c-2n} \rangle^\frac12= 5.01\,$fm, giving
 respectively $\sqrt{\langle r^2_{nn}\rangle  S_{2n}}=0.80$ and
 $\sqrt{\langle r^2_{c-2n}\rangle  S_{2n}}=0.55$,  consistent with our results for Bonn-A potential, which has effective range by 10\% smaller. The fine tuning of the calculations requires all low-energy parameters to be identical. Despite of that the comparison shows the extend of model dependence that the scaling functions may have.

 \begin{figure}[t]
\begin{center}

\includegraphics[angle=0,width=0.46\textwidth]{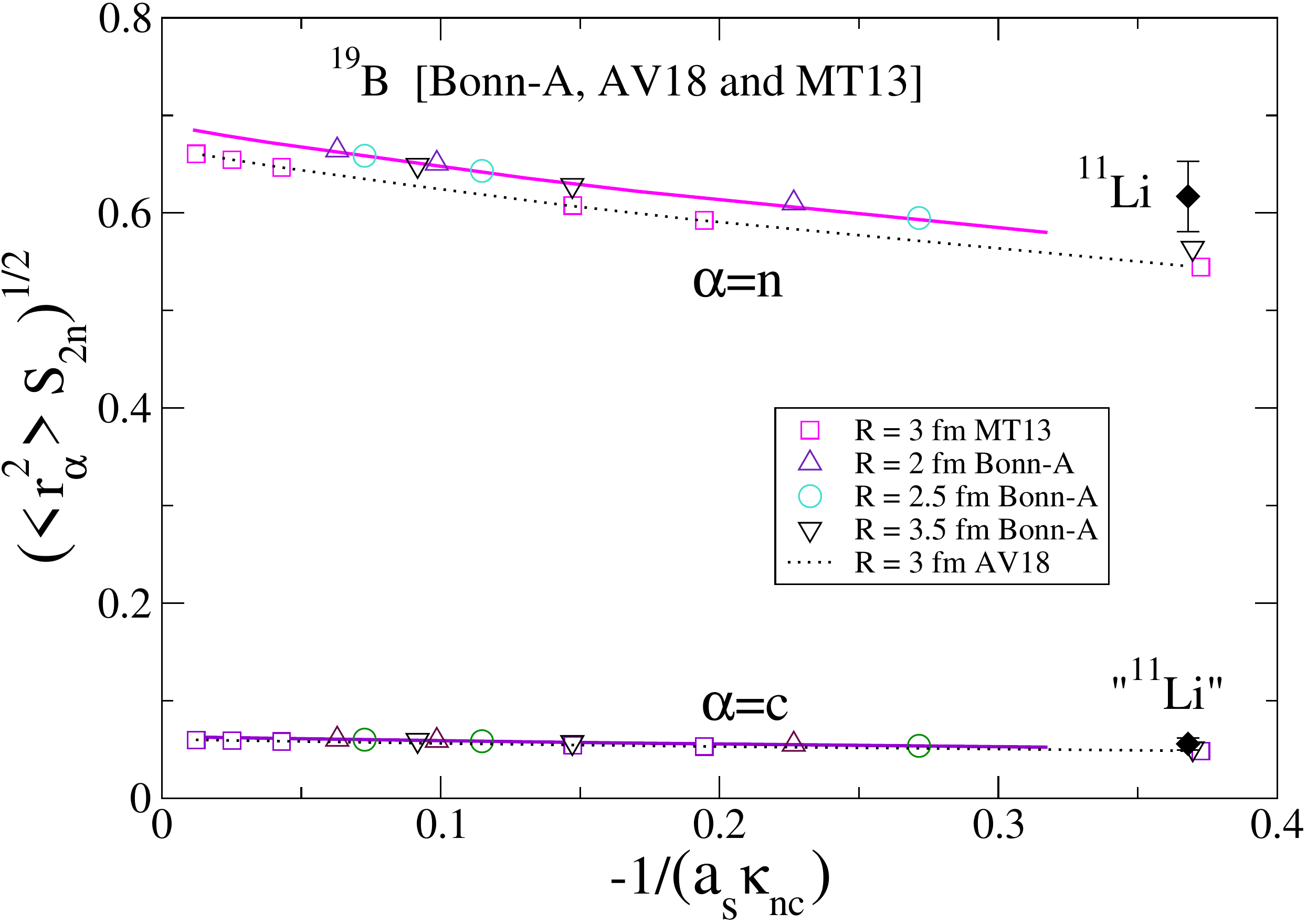}
\vspace{1cm}

\includegraphics[angle=0,width=0.46\textwidth]{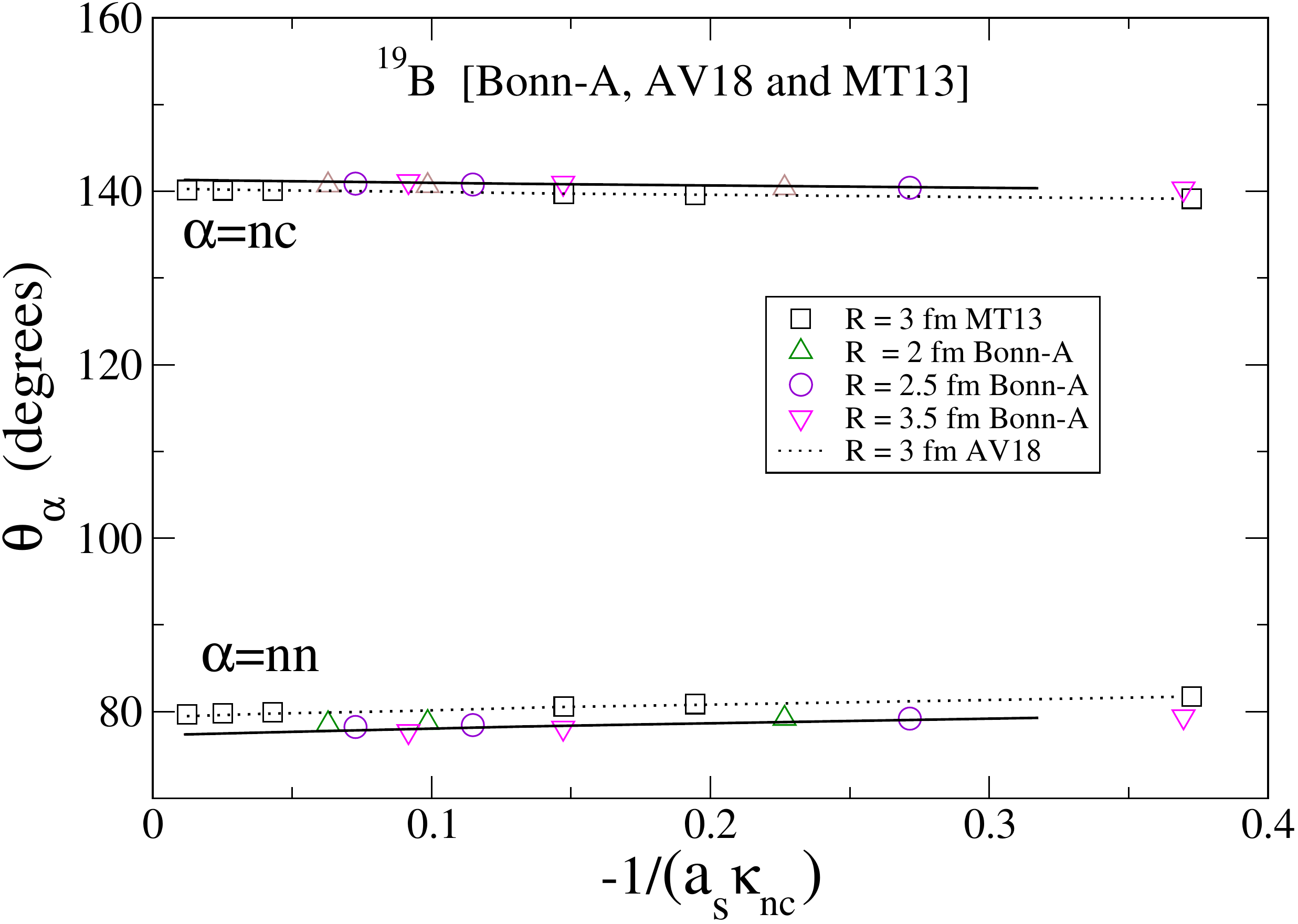}

\caption{Upper panel: $^{19}$B dimensionless products $(\langle r_\alpha^2\rangle S_{2n})^\frac12$ $(\alpha\equiv n\, , c)$    as a function of $-1/(a_s\kappa_{nc})$, where $\kappa_{nc}=(2\mu_{nc}S_{2n})^\frac12$.  Lower panel:
average angles $ \theta_\alpha $ $(\alpha\equiv nn\, , nc)$   as a function of $-1/(a_s\kappa_{nc})$.  Results for neutron-neutron Bonn-A potential (solid lines) and  MT-potential (empty squares) for $R=3\,$fm. The calculations with $R$ equals to 2, 2.5, and 3.5 fm obtained with Bonn-A potential are the up-triangle, circle and down-triangle, respectively. The results for AV18 for $R=3\,$fm are shown by the dotted line. The diamonds are the extracted values from experimental data for $^{11}$Li and "$^{11}$Li" refers to the $(2/19)(\langle ^{11}\text{Li} | r_{c-2n}^2| ^{11}\text{Li}\rangle S^{^{11}\text{Li}}_{2n})^\frac12$ (see text for the explanation). }\label{Fig:19B-Bonn+MT-2}
\end{center}
\end{figure}
 
The  data with error bar in Fig.~\ref{Fig:19B-Bonn+MT-1} is obtained from the radius extraction performed in~\cite{CookPRL2020}, where $^{19}$B has been studied by exclusive measurements of $^{17}$B+n+n from the Coulomb breakup reaction with Pb at 220 MeV/nucleon.
Three-body calculations were used to reproduce the soft E1 energy excitation spectrum below 6~MeV, and indicated  that the valence neutrons have a significant dineutron correlation in a s-wave configuration. Furthermore,
  the  $^{17}$B+n+n three-body model calculations performed to analyze the experimental results uses $a_s=-50\,$fm and $S_{2n}=0.5\,$MeV,  it was extracted the value for $\langle r^2_{c-2n}\rangle^{\frac12} $ written in~\eqref{eq:exprc2n},
which gives 
$(\langle r^2_{c-2n}\rangle  S_{2n})^\frac12=0.632\pm 0.026$.

We  also compared the results for the scaling function given in Fig.~\ref{Fig:19B-Bonn+MT-1} to   
the estimated $^{11}$Li core-2n distance, which is found within the interval  $5.01\pm0.32$~\cite{NakamuraPRL2006} to $6.2\pm0.5\,$fm~\cite{TanihataPPNP2013}, corresponding to
$$0.47\lesssim(\langle^{11}\text{Li}| r^2_{c-2n}| ^{11}\text{Li}\rangle S^{^{11}\text{Li}}_{2n})^\frac12\lesssim 0.59\, ,$$ using the experimental value  $S^{^{11}\text{Li}}_{2n}= 369.15(65)\,$keV~\cite{SmithPRL2008}. This last value gives  
$1/(a_{n-^9{\text{Li}}}\kappa_{nc})=-0.368$ considering the s-wave virtual state of $^{10}$Li to be at $-50\,$keV.

The upper panel of Fig.~\ref{Fig:19B-Bonn+MT-2} shows the $^{19}$B dimensionless products $(\langle r_\alpha^2\rangle S_{2n})^\frac12$ $(\alpha\equiv n\, , c)$   as a function of $-1/(a_s\kappa_{nc}) $ calculated with Bonn-A, MT13 and AV18 considering $R=3\,$fm in the neutron-core potential. Other values of $R$ equals to 2, 2.5 and $3.5\,$fm were also used together with Bonn-A potential. We also present the values estimated with the extracted matter radius from the experimental data considering $^{11}$Li a  two-neutron halo nucleus for $(\langle^{11}\text{Li}| r^2_n| ^{11}\text{Li}\rangle S^{^{11}\text{Li}}_{2n})^\frac12=0.617(36)$~\cite{Fred2012} and 
$(2/19)(\langle^{11}\text{Li}| r^2_{c-2n}| ^{11}\text{Li}\rangle S^{^{11}\text{Li}}_{2n})^\frac12$, which is in the range of 0.049 and 0.056 and denoted by "$^{11}\text{Li}$" in the figure.

The lower panel of Fig.~\ref{Fig:19B-Bonn+MT-2} shows the $^{19}$B average angles $ \theta_\alpha $ $(\alpha\equiv nn\, , nc)$   as a function of $-1/(a_s\kappa_{nc}) $.  Results are displayed for the neutron-neutron Bonn-A potential considering $R$ with values of 2, 2.5, 3 and 3.5~fm. For MT13 and AV18 with  the neutron-core potential has $R=3\,$fm. As seen the angles are sensitive to the 10\% difference in $a_{nn}$  between Bonn-A and AV18 (or MT13). Interesting to observe that towards unitarity the relative angle between the neutrons decreases, as observed by comparing the Bonn-A and AV18 (or MT13) results. This comes from the swelling of the halo  from the Borromean situation to the unitarity limit, when $1/(a_s\kappa_{nc}) $ is fixed.

\section{Final Remarks}
\label{sect:finalremarks}

We conclude from the above results, that the two-neutron halo structure properties are remarkably model independent, and driven by the two-neutron separation energy and scattering lengths as long as the size of the $^{17}$Be is substantially smaller than the halo size. 
Indeed, our model confirmed that the range of  the $n-^{17}$Be potential $\mu^{-1}\sim 1/0.7\,$fm is  appreciably smaller than the halo size. The different correlations of dimensionless quantities built for the different radii, two-neutron separation energy and scattering lengths express limit cycles governed by the Efimov-like behavior of the shallow halo ground-state, and the proximity to unitarity. Noteworthy, 
our numerical results for the associated  scaling functions are quite independent on the  radius parameter around $R\sim 3\,$fm, which regulates the region where Pauli exclusion principle turns the neutron-core interaction repulsive.

Assuming that the $^{17}$B core satisfies the condition of being considerably smaller than the neutron halo,  we have, actually, two  quantities that are not well known: the two-neutron separation energy in $^{19}$B and the n-$^{17}$Be scattering length (in the hypothesis that one can disregard the spin-spin interaction leading to the  independence  on the spin states $S=1$ and $S=$2 states). 

Considering the recent data  for $^{19}$B radius
extracted from the collision experiment presented in Ref.~\cite{CookPRL2020} for
$(\langle r^2_{c-2n}\rangle)^\frac12$ given in~\eqref{eq:exprc2n} and assuming the bounds of
\begin{equation}
 0<-1/[a_s(2\mu_{nc}S_{2n})^\frac12]\lesssim 0.4\, , \nonumber
 \end{equation}
 which gives the range
 \begin{equation}
 0.5\lesssim(\langle r_\alpha^2\rangle S_{2n})^\frac12\lesssim 0.6\, ,
 \nonumber
 \end{equation}
 leading to the estimation
 $0.3\lesssim S_{2n}\lesssim 0.45\,$MeV. If one takes into account the scaling function for AV18 fitted as:
 %y=0.56825434-0.3009283*x
 %y = 0.573761 -0.419103*x+ 0.318816*x^2
\begin{equation}
(\langle r^2_{c-2n}\rangle S_{2n})^{\frac12}= 
0.5738+\frac{0.4191}{a_s\kappa_{nc}}+\frac{0.3188}{(a_s\kappa_{nc})^2}\dots\, ,
\end{equation}
one finds  $S_{2n}\simeq  0.384\pm0.036
\,$MeV obtained from   the extracted value~\cite{CookPRL2020} of $ \langle r^2_{c-2n}\rangle^\frac12$ given in \eqref{eq:exprc2n} and $a_s=-150\,$fm 
with a 50\% error, consistent with the experimental upper bound~\cite{MSU_2012}. Note that this  value of $S_{2n}$ is well within the estimation  given above and the experimental range~\cite{B19_mass,WangChin2017}.

\begin{figure}[thb!]
\begin{center}

\includegraphics[angle=0,width=0.30\textwidth]{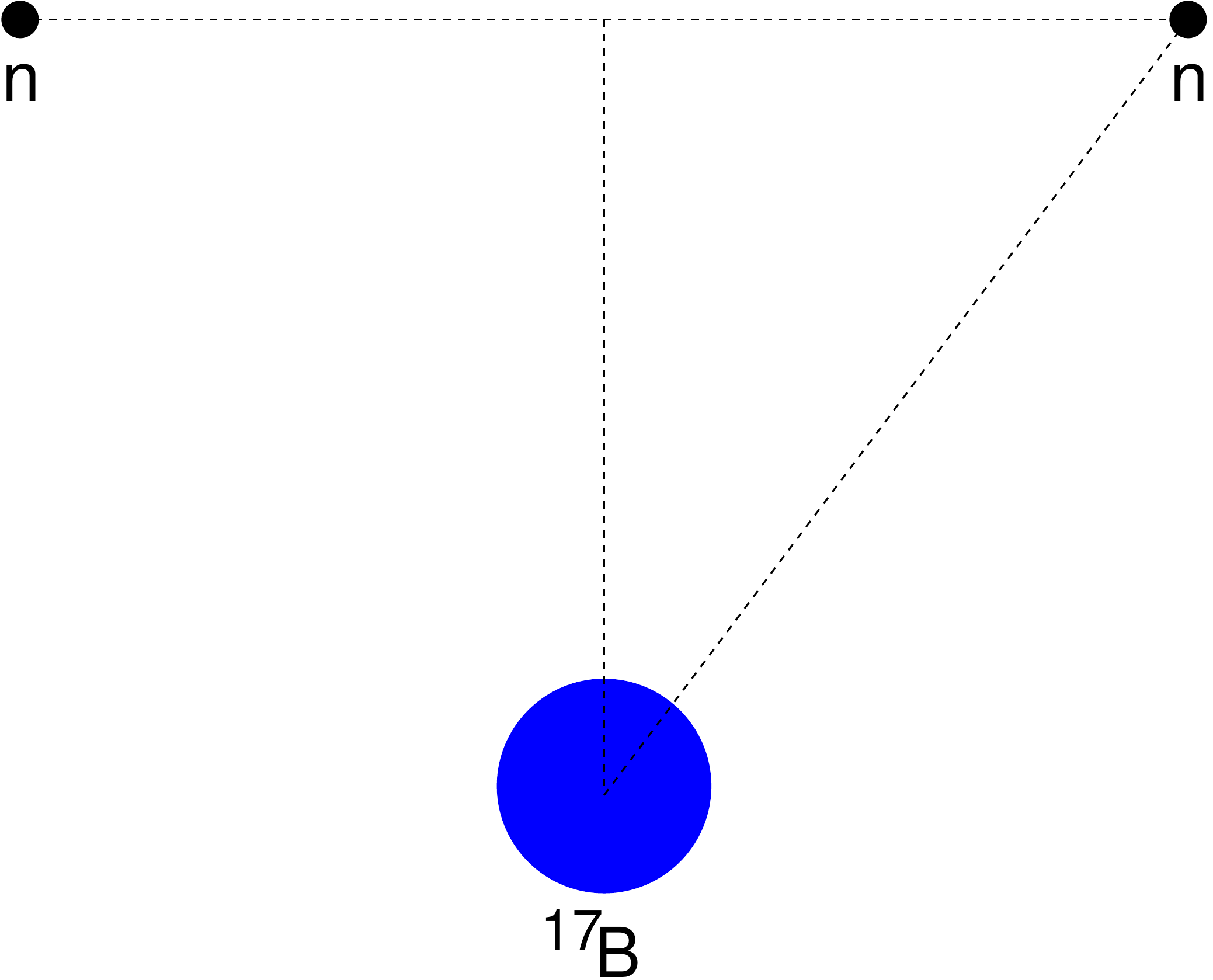}

\caption{ Geometry of $^{19}$B according to the scaled rms radii of Fig 5. Results correspond to AV18 and to the inverse scaled scattering length  $[1/a_s\kappa_{nc}]$=-0.1325.}\label{Geometry_B_n_n}
\end{center}
\end{figure}

The parametrizations of the scaling functions in the upper panel of Fig.~\ref{Fig:19B-Bonn+MT-2} are given by:
\begin{equation}
 (\langle  r^2_n\rangle S_{2n})^\frac12= 0.6655+\frac{0.4426}{a_s\kappa_{nc}}+\frac{0.3175}{(a_s\kappa_{nc})^2}\dots\, ,\label{eq:rnfit}
%0.66550254)+(-0.44260395)*x+(0.31745292)*x*x
%rn y=(0.66001932)+(-0.32493441)*x
\end{equation}
and
\begin{equation}
(\langle  r^2_c\rangle S_{2n})^\frac12=0.06039+
\frac{0.04440}{a_s\kappa_{nc}}+\frac{0.03418}{(a_s\kappa_{nc})^2}\dots\, , \label{eq:rcfit}
%=(0.060390268)+(-0.04440181)*x+(0.03417978)*x*x
% rc y=(0.059799896)+(-0.03173247)*x
\end{equation}
in units such that $\hbar=m_n=1$. 

 The $^{17}$B proton and matter radii were extracted in~\cite{Estrade:2014aba} from the experimental charge-changing cross-sections of secondary beams obtained in the FRS, GSI and Darmstadt facilities. The results from Table I in that reference are: $r^\text{ex}_p=2.67(2)\,$fm, $r^\text{ex,scaled}_p=2.63 (7)\,$fm and $r_m^\text{ex} = 3.00 (6)$, respectively. This allows us to obtain both the proton  and matter radius of $^{19}$B, by taking into account the previous estimations for $S_{2n}$ and $a_s$ introduced in the fit functions of $\langle r_n^2\rangle^\frac12$, Eq.~\eqref{eq:rnfit}, and 
 $\langle r_c^2\rangle^\frac12$,Eq.~\eqref{eq:rcfit}, together with:
 $$r^{^{19}\text{B}}_m= \sqrt{\frac{17}{19}[r_m^{ex}]^2+\frac{2}{19}\langle  r^2_n\rangle}~\text{and}~ r^{^{19}\text{B}}_p= \sqrt{[r_p^{ex}]^2+\langle  r^2_c\rangle}\, ,$$
which gives:
\begin{equation}
\begin{aligned}
&r^{^{19}\text{B}}_m=3.57\pm0.08\,\text{fm}, 
 \\
&r^{^{19}\text{B}}_p=2.74\pm0.02\,\text{fm}~\text{for}~ (r_p^\text{ex})\, ,
 \\
&r^{^{19}\text{B}}_p=2.70\pm0.07\,\text{fm} ~\text{for}~ (r_p^\text{ex,scaled})\, . 
 \end{aligned}
 \label{eq:rmrp}
\end{equation}

Finally,  in Fig.~\ref{Geometry_B_n_n} we just illustrate a geometrical
representation of  $^{19}$B nucleus, taking for distance the scaled rms radii of Fig.~\ref{Fig:19B-Bonn+MT-1}
at the inverse scaled scattering length  $1/(a_s\kappa_{nc}) $=-0.1325, quoted as  the experimentally extracted value~\cite{CookPRL2020}, and using our estimated two-neutron separation energy of $0.384\,$MeV it results in $a_s= -57\,$fm,  close to the lower bound of our suggested range of values for $a_s=-150\pm75\,$fm.  The  geometric picture gives for the angle with vertex centered at the $^{17}$B nucleus a value of $ 37^{\circ}$, close to our calculation of $\theta_{nn}/2\simeq 40^{\circ}$. Furthermore
 from Fig.~\ref{Fig:19B-Bonn+MT-1}, we have at $1/(a_s\kappa_{nc}) $=-0.1325 the values
$(\langle r^2_{nn}\rangle S_{2n})^\frac12=0.789$, $(\langle r^2_{nc}\rangle S_{2n})^\frac12=0.654$ and $(\langle r^2_{c-2n}\rangle S_{2n})^\frac12=0.524$, resulting in the values of the rms relative separation distance between the neutrons,  the neutron-core  and core-2n, respectively, of $\langle r^2_{nn}\rangle^\frac12= 8.2\,$fm,
$\langle r^2_{nc}\rangle^\frac12= 6.8\,$fm, and  $\langle r^2_{c-2n}\rangle^\frac12= 5.4\,$fm. The latter is within the error of the  extracted value~\cite{CookPRL2020}, as written in Eq.~\eqref{eq:exprc2n}. The  matter and proton radius become: $r^{^{19}\text{B}}_m=3.5\,\text{fm}~  \text{and}~ r^{^{19}\text{B}}_p= 2.7\,\text{fm}\,. $
These values can be compared to the estimations given in~\eqref{eq:rmrp}. 

The difference  between our estimations reflects the disagreement  between the results from Ref.~\cite{CookPRL2020} and our calculations shown in Fig.~\ref{Fig:19B-Bonn+MT-1}, that does not allow to narrow the knowledge on both the two-neutron separation energy  and the $^{17}$B+n scattering length, which calls for further experimental data and analysis. 
Despite of that,  our study of the universal properties of the two-neutron halo of $^{19}$B endorses the presence of a long-range s-wave correlation between the two neutrons, as well as, among the neutron- and the core, exceeding by far the ranges of the nn and n-core interactions, which gives further support to the model independence of the present findings.

\section*{Acknowledgments}
The authors thank Val\'erie Lapoux
for helpful discussions on the experimental charge and matter radii of $^{17}$B and  $^{19}$B.
We were granted access to the HPC resources of TGCC/IDRIS under the allocation A0110506006 made by GENCI (Grand Equipement National de Calcul Intensif).
This work was supported by French IN2P3 for a theory project ``Neutron-rich light unstable nuclei'',
 by the Japanese  Grant-in-Aid for Scientific Research on Innovative Areas (No.18H05407), by
 CNPq  grant 308486/2015-3,  INCT-FNA project 464898/2014-5 and  FAPESP Thematic grants 2017/05660-0 and 2019/07767-1.
This research, initiated during the  program
Living Near Unitarity at the Kavli Institute for Theoretical
Physics (KITP),  University of Santa Barbara (California), was supported in part by the National Science Foundation under Grant No. NSF PHY-1748958.

\end{document}